\documentclass{PoS}
\usepackage{graphicx,cite}
\usepackage{epsfig,bm,amsmath}

\usepackage{color}

\usepackage[T1]{fontenc} % if needed
\newcommand{\pa}{\mbox{\scriptsize\sf a}}    % PARTICELLA A
\newcommand{\pb}{\mbox{\scriptsize\sf b}}    % PARTICELLA B
\newcommand{\pq}{\mbox{\scriptsize\sf q}}    % PARTICELLA QUARK
\newcommand{\as}{\alpha_s}                   % COSTANTE FORTE 
\newcommand{\ab}{\overline{\alpha}_s}
\newcommand{\kk}{{\bf k}}            % VETTORI TRASVERSI
\newcommand{\ku}{{\bf k}_1}
\newcommand{\kd}{{\bf k}_2}
\newcommand{\qq}{{\bf q}}
\newcommand{\dif}{{\rm d}}                   % DIFFERENZIALI
\newcommand{\du}{\dif[\ku]}\newcommand{\dd}{\dif[\kd]}

\newcommand{\G}{{\cal G}}                    % FUNZIONE DI GREEN
\newcommand{\K}{{\cal K}}                    % QUANTITA' UNIVERSALE
\newcommand{\e}{\varepsilon}                 % DIMENSIONI EXTRA
\renewcommand{\l}{\lambda}
\renewcommand{\o}{\omega}
\newcommand{\g}{\gamma}                   % \gamma
\newcommand{\beq}{\begin{equation}}
\newcommand{\eeq}{\end{equation}}
\newcommand{\bea}{\begin{eqnarray}}
\newcommand{\eea}{\end{eqnarray}}
\newcommand{\non}{\nonumber}

\title{Heavy quark impact factor and single bottom quark 
production at the LHC}

\ShortTitle{Heavy quark impact factor}

\author{Grigorios Chachamis \\
Instituto de F\'{\i}sica Corpuscular, Universitat de Val\`encia -- 
Consejo Superior de Investigaciones Cient\'{\i}ficas, 
Parc Cient\'{\i}fic, E-46980 Paterna (Valencia), Spain \\
E-mail: \email{grigorios.chachamis@ific.uv.es}}

\author{\speaker{Michal De\'ak} \\
Instituto de F\'{\i}sica Corpuscular, Universitat de Val\`encia -- 
Consejo Superior de Investigaciones Cient\'{\i}ficas, 
Parc Cient\'{\i}fic, E-46980 Paterna (Valencia), Spain \\
E-mail: \email{michal.deak@ific.uv.es}}

\author{Germ\'an Rodrigo \\
Instituto de F\'{\i}sica Corpuscular, Universitat de Val\`encia -- 
Consejo Superior de Investigaciones Cient\'{\i}ficas, 
Parc Cient\'{\i}fic, E-46980 Paterna (Valencia), Spain \\
E-mail: \email{german.rodrigo@csic.es}}

\abstract{
We calculate the finite part of the heavy quark impact factor at next-to-leading 
logarithmic accuracy in a form suitable for phenomenological 
studies such as the calculation of the cross-section for
single bottom quark production at the LHC. }

\FullConference{XXI International Workshop on 
Deep-Inelastic Scattering and Related Subjects - DIS2013,\\
22-26 April 2013 \\
Marseille, France}

\begin{document}

\maketitle

\flushbottom

\section{Introduction}

Major developments in the last two decades in 
small-$x$ physics made possible the
phenomenological 
analysis of deep inelastic scattering (DIS) processes 
within the $k_T$ factorization scheme. 
They were mainly driven by the Balitsky-Fadin-Kuraev-Lipatov (BFKL) framework
for the resummation of  high center-of-mass energy logarithms
at leading~\cite{BFKLLO} and next-to-leading~\cite{BFKLNLO}
logarithmic accuracy.

Indeed, a lot of
studies were dedicated on setting up the stage for studying DIS
processes within the BFKL formalism~\cite{Salam:1998tj} and recently there were 
successful attempts for the
detailed description of the $Q^2$ and $x$ dependence of the structure functions 
$F_2$ and $F_L$  by making use of a collinearly-improved BFKL equation at 
next-to-leading logarithmic accuracy~\cite{Hentschinski:2013id}.

Apart from the gluon density, another
key ingredients for studying DIS processes
are the impact factors which are process dependent objects.
The impact factors for gluons and  massless quarks have been
calculated in Ref.~\cite{Ciafaloni:1998hu}, at next-to-leading logarithmic
accuracy (NL$x$).
This allows for the calculation of various 
DIS and `double DIS' processes with massless quarks and gluons in the initial state. 
The generalization to hadron-hadron collisions has also  
been established~\cite{Mueller:1986ey,Vera:2007kn,Kwiecinski:2001nh}.

%Marquet:2007xx,Aurenche:2008dn

The NL$x$ impact factor for a massive quark in the initial state has been calculated
in Ref.~\cite{Ciafaloni:2000sq}. However, the result was
written in the form of a sum of an infinite number of terms. 
To make that result of Ref.~\cite{Ciafaloni:2000sq} available for
phenomenological studies we recalculate the next-to-leading order heavy quark 
impact factor in a compact and resummed form which is more suitable 
for numerical applications.

\section{$k_T$-factorization}

We start with some useful definitions. The differential cross-section 
of the high-energy scattering of two partons, $\pa$ and $\pb$, can be written in a 
factorized form in terms of the gluon Green's function $\G_{\o}$ 
and the impact factors of the two partons, $h_{\pa}$ and $h_{\pb}$
respectively: 
\begin{equation}
 \frac{\dif\sigma_{\pa\pb}}{\du\,\dd}=
 \int\frac{\dif\o}{2\pi i\o} \, h_{\pa}(\ku) \, \G_{\o}(\ku,\kd) \, 
 h_{\pb}(\kd) \, \left(\frac{s}{s_0(\ku,\kd)}\right)^{\o}~.
\label{fatt}
\end{equation}
We adopt $\dif[\kk]=\dif^{2+2\e}\kk/\pi^{1+\e}$ as
the transverse phase-space measure.
The leading $\log x$ impact factor,\, $h_{}^{(0)}$, can be written as
\begin{equation}
 h_{}^{(0)}({\kk})=\sqrt{\frac{\pi}{N_c^2-1}}\;
 \frac{2C_F \as N_{\e}}{\kk^2\,\mu^{2\e}}~, \quad \text{where}\quad
 N_{\e} = \frac{(4\pi)^{\e/2}}{\Gamma(1-\e)}~,
\label{hzero}
\end{equation}
and has the same form for quarks and gluons, whereas
$\mu$ is the renormalization scale, and 
\begin{equation}\label{eq:alphas}
\ab = \frac{\as N_c}{\pi}~, \qquad
\as = \frac{g^2\Gamma(1-\e)\mu^{2\e}}{(4\pi)^{1+\e}}~,
\end{equation}
is the dimensionless strong coupling constant. We also define
\begin{equation}\label{eq:Aep}
A_{\e} =  \kk^2 \, h_{}^{(0)}({\kk}) \,
 \frac{\ab}{\Gamma(1-\e)\mu^{2\e}}~,
\end{equation}
to factor out the dependence on the strong coupling constant
and color factors.

\section{The NLO heavy quark impact factor}

According to Ref.~\cite{Ciafaloni:2000sq}, 
the NL$x$ correction to the heavy quark impact factor can be written as 
a sum of three contributions:
\begin{equation}
h_{\pq}^{(1)}(\kd) = h_{\pq,m=0}^{(1)}(\kd) 
+ \int_0^1 \dif z_1 \int \du \Delta F_{\pq}(z_1,\ku,\kd)  
+ \int \du \, \ab \, h_{\pq}^{(0)}(\ku) \, K_0(\ku,\kd) \,
\log \frac{m}{k_1} \, \Theta_{m \, k_1}~,
\label{eq:h1mass}
\end{equation}
where $k_1=|\ku|$, $\Theta_{m \, k_1} = \theta(m-k_1)$ and
\begin{equation}
\ab \, K_0(\ku,\kd) = 
\frac{\ab}{\qq^2 \Gamma(1-\e)\mu^{2\e}}
+ 2 \o^{(1)}(\ku^2) \delta[\qq]~, \,\, \quad \text{where}\quad
\delta[\qq]=\pi^{1+\e} \delta^{2+2\e}(\qq)~,
\end{equation}
is the leading order BFKL kernel. $\qq=\ku+\kd$ and
\begin{equation}\label{eq:omega}
 \o^{(1)}(\kk^2)=-\frac{g^2 N_c \kk^2}{(4\pi)^{2+\e}} \,
 \int\frac{\dif[{\bf p}]}{{\bf p}^2(\kk-{\bf p})^2}=
 -\frac{\ab}{2\e}\frac{\Gamma^2(1+\e)}{\Gamma(1+2\e)}
 \left(\frac{\kk^2}{\mu^2}\right)^{\e}~,
\end{equation}
is the gluon Regge trajectory.
The first term on the right-hand side of Eq.~\eqref{eq:h1mass} 
is the massless NL$x$ correction to the impact factor
which was computed in Ref.~\cite{Ciafaloni:1998hu}. 
Here, we discuss the last two terms in the right-hand side of Eq.~\eqref{eq:h1mass}.

\subsection{The $\Delta F_{\pq}$ term}\label{sec:dFq}

The second term in the right-hand side of Eq.~\eqref{eq:h1mass}
reads in momentum space:
\begin{align}
\Delta F_{\pq}(\kd) & = \Delta F_{\pq,real}(\kd)+\Delta F_{\pq,virt}(\kd) \non \\
& = A_{\e} \Bigg[
 \frac{\Gamma(-\e)}{2(1+2\e)} \frac{(m^2)^{\e}}{\kd^2} 
+ \frac{\Gamma(1-\e)}{2} \bigg\{
\int_0^1 \int_0^1 \dif z_1 \, \dif x 
\left(\frac{1-z_1}{z_1} +\frac{1+\e}{2}z_1 \right) \non \\
& \qquad \times \bigg[ \frac{1}{\left[x(1-x)\kd^2+m^2z_1^2\right]^{1-\e}}
-\frac{1}{\left[x(1-x)\kd^2\right]^{1-\e}} \bigg]  \non \\
& \qquad + \frac{2m^2}{\kd^2} \int_0^1 \int_0^1
\frac{ z_1(1-z_1) \, \dif z_1 \, \dif x}
{\left[x(1-x)\kd^2+m^2z_1^2\right]^{1-\e}} \bigg\}  \Bigg]~.
\label{realvirtual}
\end{align}
We cannot calculate the integral directly but we can calculate 
its Mellin transform,
\begin{equation}\label{eq:impfnlo}
\Delta \tilde{F}_{\pq}(\gamma,\e) =  
\Gamma(1+\e) \, (m^2)^{-\e} \int \dd  \left( \frac{\kd^2}{m^2}\right)^{\g-1}
\Delta F_{\pq}(\kd)~,
\end{equation}
to get finally in $\gamma$-space:
\begin{align}\label{eq:Melltr}
\Delta \tilde{F}_{\pq}(\gamma,\e) & = A_{\e} \, (m^2)^{\e} \, 
\frac{\Gamma(\g+\e)\Gamma(1-\g-2\e)\Gamma^2(1-\g-\e)}
{8\Gamma(2-2\g-2\e)} \non \\ & \times  
\bigg[ \frac{1+\e}{\g+2\e} + \frac{2}{1-2\g-4\e}
\left( \frac{1}{1-\g-2\e}- \frac{1}{3-2\g-2\e} \right) \bigg]~.
\end{align}

In Ref.~\cite{Ciafaloni:2000sq}, the residua of Eq.~\eqref{eq:Melltr} at 
the poles $\g=1-\e$ and $\g=1-2\e$ 
were studied and the singular terms in $\e$ were isolated. For the remaining poles, 
the limit $\e\rightarrow 0$ was taken and the residua were summed. 
It turns out, however, that the resulting infinite sums\footnote{See Ref.~\cite{Ciafaloni:2000sq}}
 are not suitable for further studies. 
Namely, the sum of residua to the left of the real axis 
for $\g>1$ does not converge in the region $4m^2/{\kk}_2^2>1$, which makes 
it impossible to obtain values of the impact factor in that region. 
We find that this obstacle can be overcome by keeping the full $\e$-dependence 
in the resummation of the pole-contributions
and isolating the singular and finite terms after resumming the infinite sum.
In the following, we keep the full $\e$-dependence, we perform the
inverse Mellin transform by summing the residua of the poles appearing for
$\gamma \ge 1-\e$ and $\gamma \le 1-2\e$ and then we expand in $\e$ to resolve 
$1/\e^2$ , $1/\e$ and finite terms.

The inverse Mellin transform is given by
\begin{equation}\label{eq:invMell} 
\Delta F_{\pq}(\kd) =
\frac{1}{m^2} \int_{1-2\e < {\rm Re}\, \g <1-\e} \frac{\dif \g}{2\pi i}  
\left(\frac{\kd^2}{m^2}\right)^{-\g-\e}
\Delta \tilde{F}_{\pq}(\gamma,\e)~. 
\end{equation}
The integral in Eq.~\eqref{eq:invMell} can be calculated by either closing the integration 
contour at infinity to the left of the pole at $\g=1-2\e$ or to the right of the
pole at $\g=1-\e$ and summing the residua within the contour. The respective
contributions are given by
$h_1^-$ and $h_1^+$ (for convenience we define $R=\frac{\kd^2}{4m^2}$):
\begin{equation}
h_1^-\left(R\right) = \sum\limits_{\gamma\le 1-2\e}{\rm Res}
\bigg\{(4R)^{1-\g-\e}\Delta\tilde{F}_{\pq}(\gamma,\e)\bigg\},
\end{equation}
\begin{equation}
h_1^+\left(R\right) = - \sum\limits_{\gamma\ge 1-\e}{\rm Res}
\bigg\{(4R)^{1-\g-\e}\Delta\tilde{F}_{\pq}(\gamma,\e)\bigg\}.
\end{equation} 
We choose to close the contour to the left and resum all the pole contributions.
These are distinct contributions from the poles located at $\g=1-2 \e$, 
$\frac{1}{2} - 2 \e$, $ - \e$, $ - 2 \e$ and finally from the poles
at $\g=-n + \e$ with $n$ being positive integer. 

\subsection{The $ K_0(\ku,\kd)$ term}

Let us now turn to the final ingredient in order to have the full NL$x$ heavy
quark impact factor with mass corrections.
For the real emission contribution to $K_0$ in Eq.~\eqref{eq:h1mass} 
we define the integral
\begin{equation}\label{eq:Im}
  I_m = \int \du \frac{\ab  h_{\pq}^{(0)}({\ku})}{\qq^2 \Gamma(1-\e)\mu^{2\e}}
  \log \frac{m}{k_1} \, \Theta_{m \, k_1}~. 
\end{equation}
Eq.~\eqref{eq:Im} can then be rewritten as:
\begin{equation}
\begin{split}\label{eq:Im2}
I_m & = \frac{A_{\e}}{2} \, \lim_{\alpha \to 0^+}
\int_{-i\, \infty}^{+i\, \infty} \frac{\dif\l}{2\pi \, i} \, \frac{1}{(\l+\alpha)^2} \, 
(m^2)^{\l} \int \frac{\du}{\qq^2 \, (\ku^2)^{1+\l}} \\
& = \frac{A_{\e}}{2} \, \lim_{\alpha \to 0^+}
\int_{-i\, \infty}^{+i\, \infty} \frac{\dif\l}{2\pi \, i} \, \frac{1}{(\l+\alpha)^2} \, 
\frac{\Gamma(1+\l-\e) \Gamma(\e) \Gamma(\e-\l)}
{\Gamma(1+\l) \Gamma(2\e-\l)} (m^2)^{\l} (\kd^2)^{-1-\l+\e}~,
\end{split}
\end{equation}
where $\alpha>0$. To recover $I_m$ the limit $\alpha\rightarrow 0$ has to be taken.
Eq.~\eqref{eq:Im2} has a form similar to Eq.~\eqref{eq:invMell}. A similar 
procedure to the one in Section~\ref{sec:dFq} can be applied in order to calculate the 
integral on the right-hand side of Eq.~\eqref{eq:invMell}. The leading poles 
containing the singular terms in $\e$ are now at $\l=-\alpha, \,\,(\alpha \rightarrow 0^{+})$, 
if the integration contour is closed to the left and at $\l=\e$ if it is closed to the right.

In analogy with $h_1^\pm\left(R\right)$ we define $h_2^\pm\left(R\right)$. After 
summing the residues of the integrand of Eq.~\eqref{eq:invMell} and expanding 
in $\e$ we obtain for $\kd^2<m^2$:
\begin{equation}
h_2^{-}\left(R\right)=h_{}^{(0)}({\kd})\,\o_{}^{(1)}({\kd})\,
\left(-\e^{-1}+\log(4R)-\e\,{\rm Li}_2(4R)\right)~,
\end{equation}
and for $\kd^2>m^2$: 
\begin{equation}
h_2^{+}\left(R\right)=h_{}^{(0)}({\kd})\,\o_{}^{(1)}({\kd})\,
\left[-\e^{-1}+\log(4R)-\e\,\left(\frac{1}{2}\log^2(4R)
+ {\rm Li}_2\left(\frac{1}{4R}\right)\right)\right]~.
\end{equation}
It is now noteworthy to explain why $h_2^-\left(R\right)\neq h_2^+\left(R\right)$ in contrast
to  $h_1^-\left(R\right)= h_1^+\left(R\right)$.
The original integral definition of
$I_m$ in Eq.~\eqref{eq:Im} contains the $\theta$-function $\Theta_{m\,k_1}$. After rewriting
it into the form of Eq.~\eqref{eq:Im2} we see that the $\theta$-function generates after
integration a discontinuity in the first derivative of $I_m$.

For $\kd^2<m^2$ we have to include also the virtual term of $K_0$, which leads to the 
contribution
\begin{equation}
h^+_V({\kd})=-h_{}^{(0)}({\kd})\, \o^{(1)}(\kd^2)
\log{\left(\frac{\kd^2}{m^2}\right)}\, \Theta_{m\,k_2}~.
\end{equation}

\subsection{Final compact expression for the heavy quark impact factor}

The contributions discussed in the previous sections can be put together
into a very compact formula. The NL$x$ heavy quark impact 
factor can be expressed as the sum
\begin{equation}\label{eq:h(1)}
h_{\pq}(\kd) = h_{\pq}^{(1)}(\kd)|_{\rm sing} + h_{\pq}(\kd)|_{\rm finite}~,
\end{equation}
where the singular term $h_{\pq}^{(1)}(\kd)|_{\rm sing}$ 
has been calculated in Ref.~\cite{Ciafaloni:2000sq}, 
and the finite contribution is given by 
\begin{eqnarray}
h_{\pq}(\kd)|_{\rm finite} &=& 
h^{(0)}_{\pq}(\alpha_s(\kd)) \, \Bigg\{1 + \frac{\alpha_{S}\, N_C}{2\pi} \, \Bigg[ 
\mathcal{K}-\frac{\pi^2}{6} + 1 
- \log(R_1) \left(\left(1+2R\right)\sqrt{\frac{1+R}{R}} + 2 \log ({R_1})\right) \non \\
&& \qquad - 3 \, \sqrt{R}
\left({\rm Li}_2\left(R_1\right)-{\rm Li}_2\left(-R_1\right)
+ \log(R_1) \, \log \left(\frac{1-{R_1}}{1+{R_1}}\right) \right) \non \\ 
&& \qquad + {\rm Li}_2\left(4R\right) \, \Theta_{m\, k_2} + \left(
  \frac{1}{2} \log\left(4R\right)+\frac{1}{2} \log^2\left(4R\right)
+ {\rm Li}_2 \left( \frac{1}{4R} \right) \right) \, \Theta_{k_2\,m} \Bigg] 
\Bigg\}~,
\end{eqnarray}
with $R_1 = (\sqrt{R}+\sqrt{1+R})^{-1}$, and $\K$ being
\begin{equation}
\K = \frac{67}{18}-\frac{\pi^2}{6}-\frac{5n_f}{9N_c}\;.
\end{equation}
As in Ref.~\cite{Ciafaloni:2000sq}, we have absorbed the singularities 
proportional to the beta function into the running 
of the strong coupling constant, $\alpha_s(\kd)$~\cite{Rodrigo:1993hc}.
%
%{\it check that this expression expanded agrees with (4.18) of }
%\cite{Ciafaloni:2000sq}
%
%the theta functions has been changed !!!! check 
%}

\section{Conclusions and outlook}

We discussed in some detail how, based on the results 
in Ref.~\cite{Ciafaloni:2000sq},
we obtained expressions suitable for the numerical
implementation of the heavy quark impact factor
in momentum space.
We have also showed, that the first derivative in the $t$-channel 
transverse momentum $\kd$ 
of the finite part of the heavy quark impact factor exhibits 
a discontinuity at the point where the transverse momentum 
is equal to the mass of the heavy quark, $|\kd|=m$.
We will present further results of the actual numerical implementation~\cite{hqif1}
of the heavy quark impact factor in momentum and $\gamma$-space~\cite{hqif2}
for phenomenological studies ---such as the single bottom quark 
production at the LHC--- elsewhere.

\section*{Acknowledgments}

This work has been supported by the 
Research Executive Agency (REA) of the European Union under 
the Grant Agreement number PITN-GA-2010-264564 (LHCPhenoNet),
by the Spanish Government and EU ERDF funds 
(grants FPA2007-60323, FPA2011-23778 and CSD2007-00042 
Consolider Project CPAN) and by GV (PROMETEUII/2013/007). 
GC acknowledges support from Marie Curie actions (PIEF-GA-2011-298582).
MD acknowledges support from Juan de la Cierva programme (JCI-2011-11382).


\begin{thebibliography}{10}


\bibitem{BFKLLO}  
L.~N.~Lipatov, 
Sov.\ J.\ Nucl.\ Phys.\  {\bf 23} (1976) 338;
%%CITATION = YAFIA,23,642;%%
%
E.~A.~Kuraev, L.~N.~Lipatov, V.~S.~Fadin,
Phys.\ Lett.\  B {\bf 60} (1975) 50, 
%%CITATION = PHLTA,B60,50;%%
Sov.\ Phys.\ JETP {\bf 44} (1976) 443,
%%CITATION = ZETFA,71,840;%%
Sov.\ Phys.\ JETP {\bf 45} (1977) 199;
%%CITATION = ZETFA,72,377;%%
%
Ia.~Ia.~Balitsky, L.~N.~Lipatov, 
Sov.\ J.\ Nucl.\ Phys.\  {\bf 28} (1978) 822. 
%%CITATION = YAFIA,28,1597;%%
%\cite{Fadin:1993qb}


\bibitem{BFKLNLO}
V.~S.~Fadin, L.~N.~Lipatov, Phys.\ Lett.\  B {\bf 429} (1998) 127 [hep-ph/9802290];
%%CITATION = PHLTA,B429,127;%%
%
%\bibitem{Ciafaloni:1998gs}
 M.~Ciafaloni, G.~Camici, Phys.\ Lett.\  B {\bf 430} (1998) 349 [hep-ph/9803389].
  %%CITATION = PHLTA,B430,349;%%


%\cite{Salam:1998tj}
\bibitem{Salam:1998tj} 
  G.~P.~Salam,
  %``A Resummation of large subleading corrections at small x,''
  JHEP {\bf 9807}, 019 (1998)
  [hep-ph/9806482];
  %%CITATION = HEP-PH/9806482;%% 
%  %\cite{Ciafaloni:1998iv}
%%\bibitem{Ciafaloni:1998iv} 
%  M.~Ciafaloni and D.~Colferai,
%  %``The BFKL equation at next-to-leading level and beyond,''
%  Phys.\ Lett.\ B {\bf 452}, 372 (1999)
%  [hep-ph/9812366];
%  %%CITATION = HEP-PH/9812366;%% 
%\cite{Ciafaloni:1999yw}
%\bibitem{Ciafaloni:1999yw} 
  M.~Ciafaloni, D.~Colferai and G.~P.~Salam,
  %``Renormalization group improved small x equation,''
  Phys.\ Rev.\ D {\bf 60}, 114036 (1999)
  [hep-ph/9905566].
  %%CITATION = HEP-PH/9905566;%%
 
 %\cite{Hentschinski:2013id}
\bibitem{Hentschinski:2013id}
  M.~Hentschinski, A.~S.~Vera and C.~Salas,
  %``Description of F2 and FL at small x using a collinearly-improved BFKL resummation,''
  arXiv:1301.5283 [hep-ph].
  %%CITATION = ARXIV:1301.5283;%%


%\cite{Ciafaloni:1998hu}
\bibitem{Ciafaloni:1998hu} 
  M.~Ciafaloni and D.~Colferai,
  %``K factorization and impact factors at next-to-leading level,''
  Nucl.\ Phys.\ B {\bf 538}, 187 (1999)
  [hep-ph/9806350].
  %%CITATION = HEP-PH/9806350;%%
  
  %\cite{Mueller:1986ey}
\bibitem{Mueller:1986ey}
  A.~H.~Mueller and H.~Navelet,
  %``An Inclusive Minijet Cross-Section and the Bare Pomeron in QCD,''
  Nucl.\ Phys.\ B {\bf 282} (1987) 727.
  %%CITATION = NUPHA,B282,727;%%
  %315 citations counted in INSPIRE as of 30 Jun 2013

%\cite{Vera:2007kn}
\bibitem{Vera:2007kn}
  A.~Sabio Vera and F.~Schwennsen,
  %``The Azimuthal decorrelation of jets widely separated in rapidity as a test of the BFKL kernel,''
  Nucl.\ Phys.\ B {\bf 776} (2007) 170
  [hep-ph/0702158 [HEP-PH]].
  %%CITATION = HEP-PH/0702158;%%
  %69 citations counted in INSPIRE as of 02 Jul 2013
  
%\cite{Kwiecinski:2001nh}
\bibitem{Kwiecinski:2001nh}
  J.~Kwiecinski, A.~D.~Martin, L.~Motyka and J.~Outhwaite,
  %``Azimuthal decorrelation of forward and backward jets at the Tevatron,''
  Phys.\ Lett.\ B {\bf 514} (2001) 355
  [hep-ph/0105039].
  %%CITATION = HEP-PH/0105039;%%
  %17 citations counted in INSPIRE as of 02 Jul 2013


  
%\cite{Fadin:1998py}
%\bibitem{Fadin:1998py} 
  %V.~S.~Fadin and L.~N.~Lipatov,
  %``BFKL pomeron in the next-to-leading approximation,''
  %Phys.\ Lett.\ B {\bf 429}, 127 (1998)
  %[hep-ph/9802290].
  %%CITATION = HEP-PH/9802290;%%
  
%\cite{Ciafaloni:1998gs}
%\bibitem{Ciafaloni:1998gs} 
  %M.~Ciafaloni and G.~Camici,
  %``Energy scale(s) and next-to-leading BFKL equation,''
  %Phys.\ Lett.\ B {\bf 430}, 349 (1998)
  %[hep-ph/9803389].
  %%CITATION = HEP-PH/9803389;%%

%%%%%%%%%%%%%%%%%%%%%%%%%%%%%%%%%%%%%%%%%%%%%%%%%%%%%%%%%%%%%%




%\cite{Ciafaloni:2000sq}
\bibitem{Ciafaloni:2000sq}
  M.~Ciafaloni and G.~Rodrigo,
  %``Heavy quark impact factor at next-to-leading level,''
  JHEP {\bf 0005} (2000) 042
  [hep-ph/0004033].
  %%CITATION = HEP-PH/0004033;%%

%\cite{Rodrigo:1993hc}
\bibitem{Rodrigo:1993hc}
  G.~Rodrigo and A.~Santamaria,
  %``QCD matching conditions at thresholds,''
  Phys.\ Lett.\ B {\bf 313} (1993) 441
  [hep-ph/9305305];
  %%CITATION = HEP-PH/9305305;%%
  %82 citations counted in INSPIRE as of 02 Jul 2013  
%\cite{Rodrigo:1997zd}
%\bibitem{Rodrigo:1997zd}
  G.~Rodrigo, A.~Pich and A.~Santamaria,
  %``Alpha-s (m(Z)) from tau decays with matching conditions at three loops,''
  Phys.\ Lett.\ B {\bf 424} (1998) 367
  [hep-ph/9707474].
  %%CITATION = HEP-PH/9707474;%%
  %68 citations counted in INSPIRE as of 02 Jul 2013

\bibitem{hqif1}
  G.~Chachamis, M.~Deak and G.~Rodrigo, work in progress.

\bibitem{hqif2}
  G.~Chachamis, M.~Deak, M.~Hentschinski, G.~Rodrigo, C.~Salas and  A.~S.~Vera, work in progress.

\end{thebibliography}
\end{document}